# Superconductivity from On-Chip Metallization on 2D Topological Chalcogenides

Yanyu Jia[1,*], Guo Yu[1,2], Tiancheng Song[1], Fang Yuan[3], Ayelet J Uzan[1], Yue Tang[1], Pengjie Wang[1], Ratnadwip Singha[3], Michael Onyszczak[1], Zhaoyi Joy Zheng[1,2], Kenji Watanabe[4], Takashi Taniguchi[5], Leslie M Schoop[3], Sanfeng Wu[1,*]

[1] *Department of Physics, Princeton University, Princeton, New Jersey 08544, USA*
[2] *Department of Electrical and Computer Engineering, Princeton University, Princeton, New Jersey 08544, USA*
[3] *Department of Chemistry, Princeton University, Princeton, New Jersey 08544, USA*
[4] *Research Center for Electronic and Optical Materials, National Institute for Materials Science, 1-1 Namiki, Tsukuba 305-0044, Japan*
[5] *Research Center for Materials Nanoarchitectonics, National Institute for Materials Science, 1-1 Namiki, Tsukuba 305-0044, Japan*

[*]Email: sanfengw@princeton.edu; yanyuj@princeton.edu

Two-dimensional (2D) transition metal dichalcogenides (TMDs) is a versatile class of quantum materials of interest to various fields including, e.g., nanoelectronics, optical devices, and topological and correlated quantum matter. Tailoring the electronic properties of TMDs is essential to their applications in many directions. Here, we report that a highly controllable and uniform on-chip 2D metallization process converts a class of atomically thin TMDs into robust superconductors, a property belonging to none of the starting materials. As examples, we demonstrate the introduction of superconductivity into a class of 2D air-sensitive topological TMDs, including monolayers of $T_d$-$WTe_2$, 1T'-$MoTe_2$ and 2H-$MoTe_2$, as well as their natural and twisted bilayers, metalized with an ultrathin layer of Palladium. This class of TMDs are known to exhibit intriguing topological phases ranging from topological insulator, Weyl semimetal to fractional Chern insulator. The unique, high-quality two-dimensional metallization process is based on our recent findings of the long-distance, non-Fickian in-plane mass transport and chemistry in 2D that occur at relatively low temperatures and in devices fully encapsulated with inert insulating layers. Highly compatible with existing nanofabrication techniques for van der Waals (vdW) stacks, our results offer a route to designing and engineering superconductivity and topological phases in a class of correlated 2D materials.

## I. INTRODUCTION

Introducing and designing superconductivity in non-superconducting quantum materials are often desired for engineering new phases of matter and superconducting (SC) quantum devices. A prominent example is the hope to create non-abelian anyons in artificial nanostructures [1–3]. For instance, introducing superconductivity to a topological insulator has been proposed for realizing the long-sought-after Majorana zero modes [3–5], an Ising type of anyons that can be used for demonstrating non-abelian braiding statistics and partial operations of a topological quantum bit. In more ambitious theoretical proposals combining superconductivity and fractional quantum Hall edge states, one may in principle realize distinct types of non-abelian states, such as the parafermion modes [2,6–9], which could achieve full operations of a topological quantum bit. However, many proposals require high-quality designable integration of superconductors with topological quantum materials, representing a key challenge from device engineering perspectives.

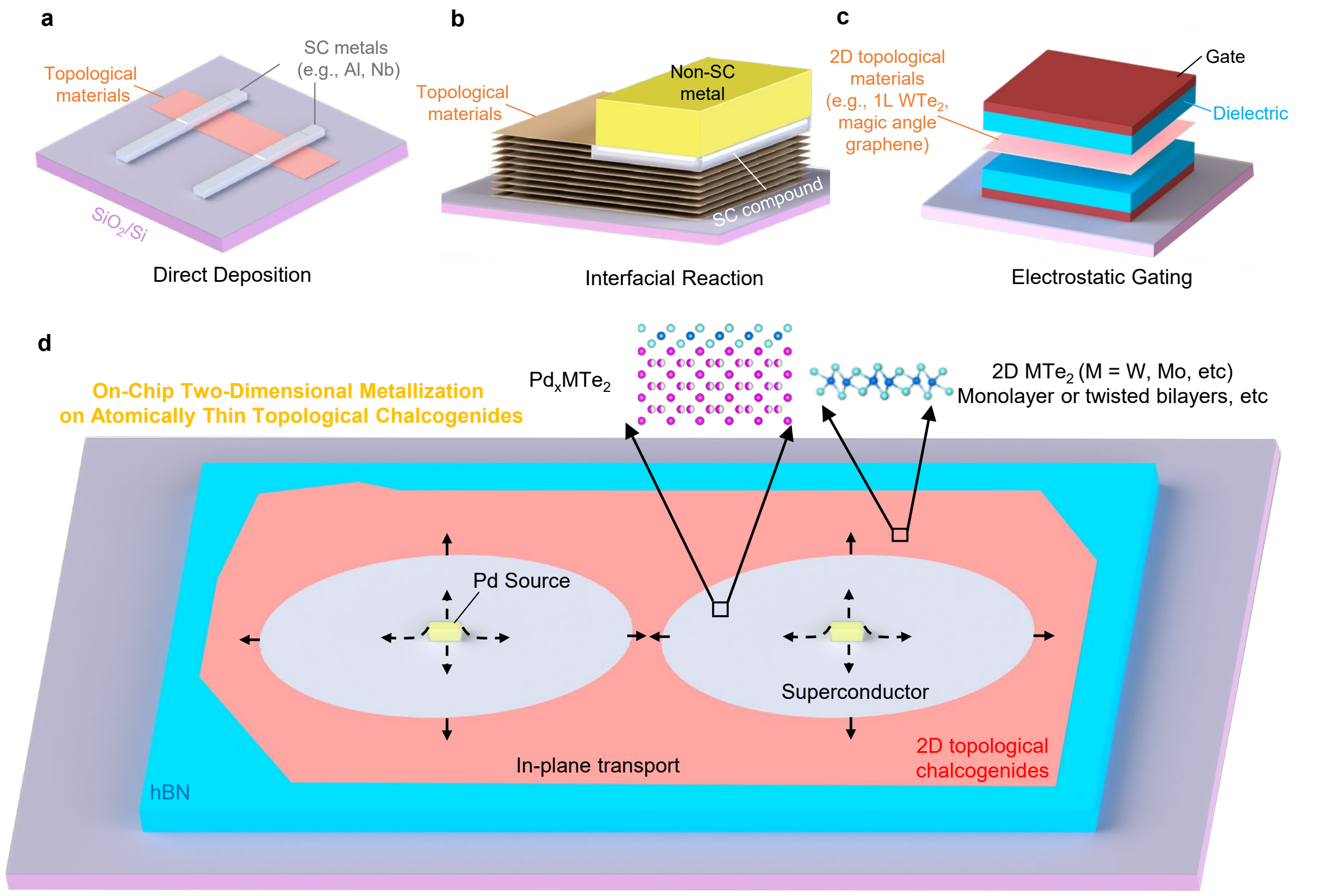

**Fig. 1. On-chip two-dimensional metallization on atomically thin topological chalcogenides. a-c.** List of known approaches for introducing superconductivity to non-superconducting materials, including direct metal deposition of metal superconductors (**a**) in a large chamber, interfacial reaction that produces superconducting compound at contact (**b**) and the rare cases of gate-induced superconductivity (**c**). **d.** Our approach of on-chip 2D metallization, based on the recent finding of the long-distance non-Fickian 2D mass transport of Pd at relatively low temperatures (ref. [27] ). The process converts regions of 2D TMDs, e.g., $MTe_2$ (M = W, Mo) into a new superconducting compound $Pd_xMTe_2$, in a controllable and designable fashion.

Conventional approaches of depositing a superconducting metal to the surface of a material (Fig. 1a) have been employed in, e.g., the nanowire [10], quantum well [11] and graphene systems [12], to name a few. However, for a class of air-sensitive 2D materials, this technique faces severe challenges in producing high quality devices. Recent studies have shown that superconducting compounds may be created at the contact vicinity between a deposited metal and a layered material due to interfacial reactions [13–17] (Fig. 1b). The superconducting contact created in this process is however nonuniform and of small volume attached to a bulk non-superconducting metal. In a study of multilayer $WTe_2$ Josephson junctions employing this method, an additional superconducting metal was deposited to improve the transport quality of the device [13]. Similar efforts have also been put forward in introducing superconductivity in epitaxy-grown topological insulators [18–21]. Superconductivity at low carrier densities may also be realized by electrostatic gating in novel 2D materials, e.g., monolayer $WTe_2$ [22,23] and magic-angle graphene systems [24–26] (Fig. 1c). However, these outstanding situations are rare and unlikely to be generalized to the diverse family of 2D quantum materials.

Here, we report the creation of robust superconductivity in a class of 2D TMDs metalized with a uniform layer of atomically thin Palladium. The results are based on our recent surprising finding [27] of a rapid, long-distance, non-Fickian (hence non-diffusive) in-plane transport of metal films on monolayer TMDs at temperatures well below the melting points of all materials involved. The process realizes on-chip 2D chemical synthesis templated on monolayer crystals (Fig. 1d), based on which we demonstrate its capability in introducing superconductivity into 2D materials. We characterize the electronic properties of the resultant new 2D compounds created in topological chalcogenides, including Pd-metalized monolayer and bilayer $T_d$-$WTe_2$, monolayer 1T'-$MoTe_2$, monolayer and twisted bilayer 2H-$MoTe_2$, and find superconductivity in all these cases.

## II. RESULTS

### A. Superconductivity in 2D Pd-metalized $WTe_2$

$WTe_2$ monolayer is an excitonic topological insulator exhibiting the quantum spin Hall effect [28–33]. Introducing superconductivity to its helical edge mode is proposed as a route to topological superconductivity and Majorana zero modes [4,5]. Superconductivity has been previously found in monolayer $WTe_2$ under electrostatic gating [22,23], in which superconducting properties are sensitive to carrier density. Here, we show a distinct approach for introducing superconductivity to monolayer and bilayer $WTe_2$ based on the on-chip 2D metallization and crystal growth method [27]. The experiments start with fabricating a vdW stack consisting of mechanically exfoliated monolayer or bilayer $WTe_2$ and hexagonal boron-nitride (hBN), placed on top of a $SiO_2$/Si substrate. Inside the stack, Pd seed islands, in contact with $WTe_2$, are pre-deposited using standard nano-lithography techniques. Upon heating the stack at ~ 200 °C, a sub-nanometer-thick layer of Pd transports from the seeds and spreads uniformly over the entire 2D flake in about an hour (Fig. 2a & b). This anomalous mass transport and the resulting new crystalline compound, with a chemical composition $Pd_7WTe_2$, were characterized in a previous work [27]. Here, we report the electronic transport properties of the new compound and find that it is a superconductor at ultralow temperatures. It is interesting to note that neither of the starting materials ($WTe_2$ and Pd) superconducts in their pristine forms.

Fig. 2c plots the four-probe resistance ($R_{xx}$) as a function of temperature ($T$) measured on $Pd_7WTe_2$ in device D1 (seeded on monolayer $WTe_2$) and D2 (seeded on bilayer $WTe_2$). D1 displays a characteristic resistance drop to zero near $T_c \sim 0.45$ K, at which $R_{xx}$ is half of its normal state value. The bilayer seeded $Pd_7WTe_2$ (D2) exhibits a higher $T_c \sim 0.9$ K. The superconducting nonlinear $IV$ curves, as well as differential resistance, are shown in Figs. 2d & e, revealing a critical current of ~ 0.4 μA in this device. A perpendicular magnetic field fully suppresses superconductivity at $B_c \sim 0.4$ T (Figs. 2f & g). These observations confirm superconductivity in this new compound. We have also fabricated a dual-gated device (D3), in which we can electrostatically vary the electron density, $n_g$, in $Pd_7WTe_2$ by $\sim \pm 1.5 \times 10^{13}$ cm$^{-2}$. Within this entire range, we find no change in the superconducting properties (Fig. 2h), implying a high carrier density in the sample. This is in sharp contrast to the gate-induced low-density superconductivity [22,23] in monolayer $WTe_2$, which develops a strong insulator state in the absence of gating induced doping ($n_g \sim 0$) [32]. Also, the critical magnetic field found in $Pd_7WTe_2$ is much larger than that found in the gate-induced superconductivity [22,23] in intrinsic monolayer $WTe_2$, further confirming a distinct origin. We further note that in our $Pd_7WTe_2$ device (D1, Fig. 2b) the data is taken when the monolayer $WTe_2$ is fully converted to $Pd_7WTe_2$, so there is no longer $WTe_2$ in the device. The atomic structure of W-Te-W in $Pd_7WTe_2$ is completely different from the pristine $WTe_2$ [27]. In Supplemental Material Fig. S1, we characterize the superconducting properties of the bilayer seeded $Pd_7WTe_2$ (D2). In Supplemental Material Fig. S2, we show a Fraunhofer-like pattern seen in the critical current measurement, induced by disorders that create an accidental junction, demonstrating the superconducting interference effects. In Supplemental Material Fig. S3, we present the measurement of the vortex Nernst effect that directly signifies the formation and motion of superconducting vortices. These comprehensive characterizations establish superconductivity in the new $Pd_7WTe_2$ compound.

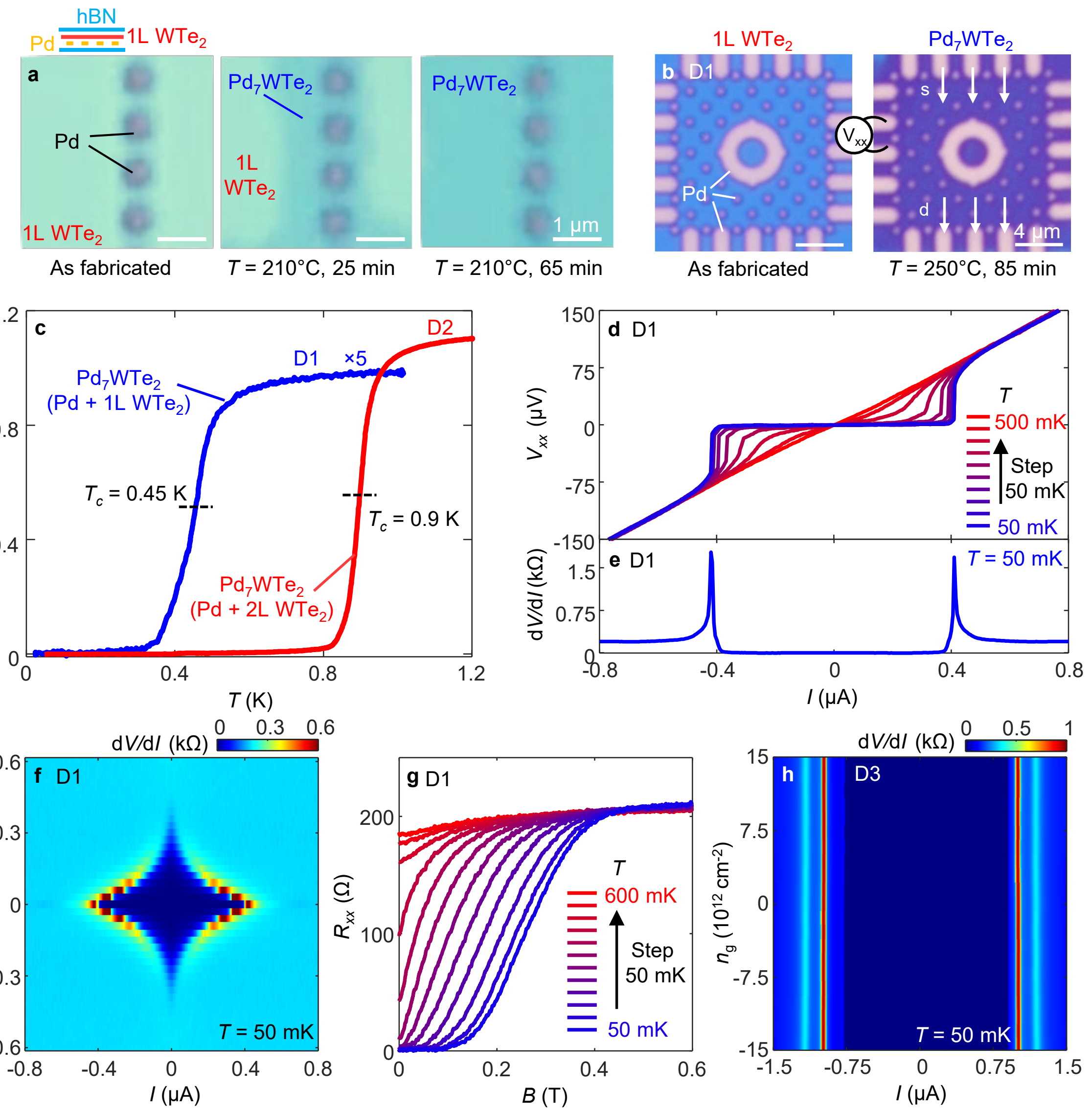

**Fig. 2. Superconductivity in $Pd_7WTe_2$. a,** Optical images of a device demonstrating the on-chip metallization process for a monolayer (1L) $WTe_2$ placed on top of Pd seeds, encapsulated by hBN. $WTe_2$ is large and covers the whole imaging window. Left: the device as fabricated. Middle: an image of the device after heat treatment at ~ 210 °C for 25 mins. Right: after 65 mins ($WTe_2$ inside the window is fully converted to $Pd_7WTe_2$). The characterization of this process and the new $Pd_7WTe_2$ compound is discussed in ref. [27]. **b,** A similar process for another device (D1), before (left image) and after (right image) the heat treatment. Here the Pd leads also serve as transport electrodes. **c,** $R_{xx}$ of $Pd_7WTe_2$ measured for D1 (blue, seeded on monolayer $WTe_2$) and D2 (red, seeded on bilayer $WTe_2$), as a function of temperature. **d,** Nonlinear *IV* curves (top, at different $T$) and, **e,** differential resistance $dV/dI$, as a function of applied DC current $I$, demonstrating the effect of critical currents. **f,** A $dV/dI$ map that reveals the vanishing critical currents under magnetic fields $B$. **g,** $R_{xx}$ as a function of $B$, taken at various $T$. **h,** $dV/dI$ *v.s.* $I$, taken over the entire gate range in D3 with gate electrodes, revealing no gate dependence.

## B. Superconductivity in 2D Pd-metalized 1T'-$MoTe_2$ and 2H-$MoTe_2$

At room temperature, $MoTe_2$ monolayer can be stabilized in two different phases, exhibiting either a monoclinic (1T') or a hexagonal lattice structure (2H). 1T'-$MoTe_2$ is known as a candidate of Weyl semimetal and develops superconductivity below 0.1 K [34,35]. In contrast, 2H-$MoTe_2$ is a semiconductor, not a superconductor. We fabricate both monolayer 1T'-$MoTe_2$ (D4) and monolayer 2H-$MoTe_2$ (D5) in contact with Pd seeds, fully encapsulated with hBN from the top and bottom. When the stack is placed at ~ 250 °C, Pd rapidly propagates in the 2D plane and reacts with the $MoTe_2$ monolayer flake, just like the Pd transportation on $WTe_2$. Figs. 3a & b display optical microscope images of the two devices (D4 & D5) before and after the heat treatment, revealing the consequence of the Pd metallization. Note that, as we have emphasized previously, the long-distance transport process here must involve chemical affinity between Pd and Te, not a simple physical diffusion. The resulting final material is a new compound $Pd_xMoTe_2$ consisting of Pd and atoms from the seed monolayers. Atomic force microscopy suggests that the thickness of the new compound is ~ 1.5 nm and the thickness increase after Pd metallization is ~ 0.8 nm (Supplemental Material Fig. S4), close to that of $Pd_7WTe_2$ obtained in $WTe_2$ case, suggesting that $x$ is close to 7 as well in the $MoTe_2$ cases. Further characterizations of the compounds are necessary to uncover their exact atomic structures in these two cases, which we leave for future study.

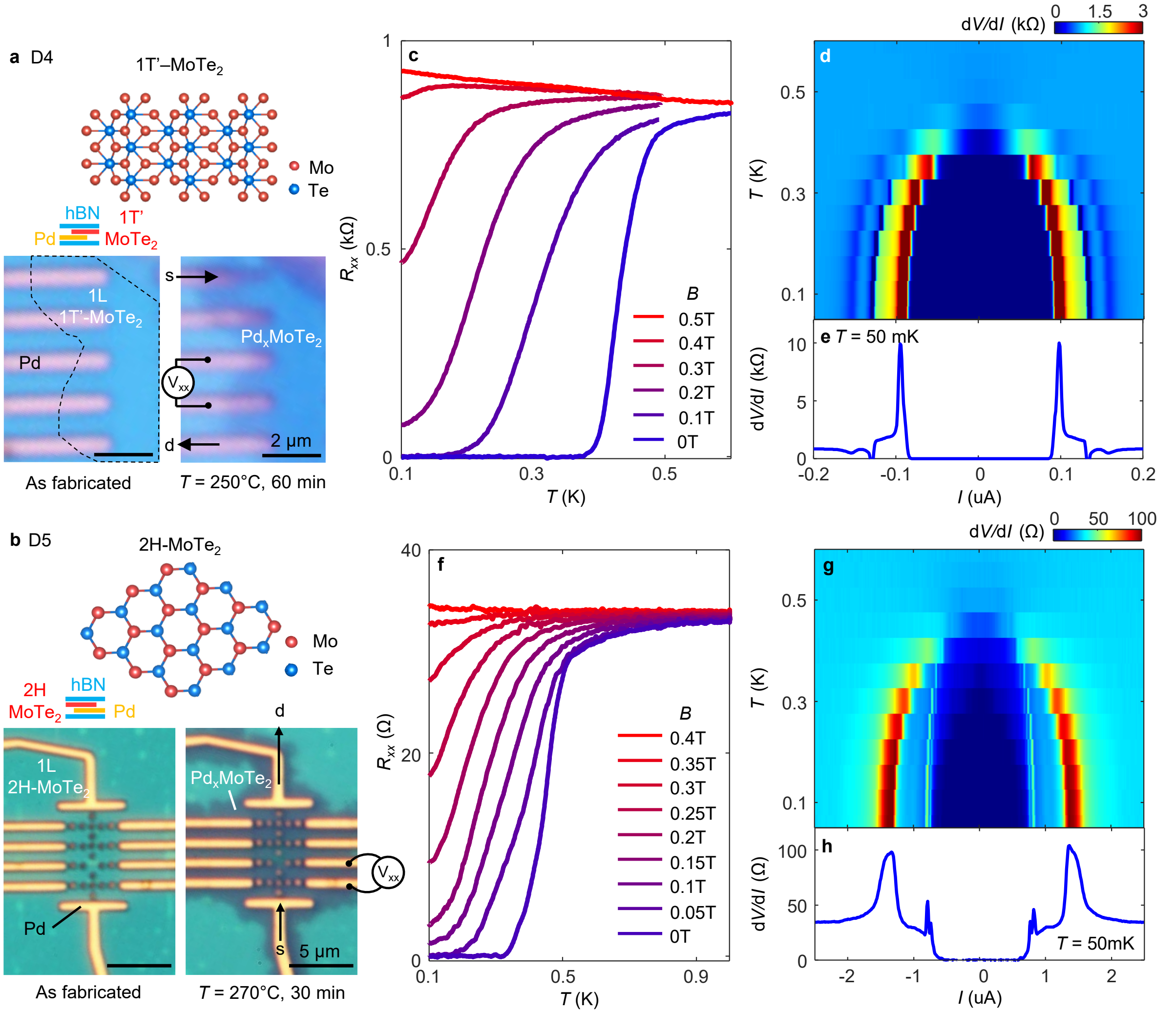

**Fig. 3. Superconductivity in $Pd_xMoTe_2$. a,** Optical images of a device (D4) for Pd metallization on monolayer 1T'-$MoTe_2$, before (left) and after (right) the heat treatment. The crystal structure of 1T'-$MoTe_2$ is shown on the top. With Pd coverage, the monolayer region become darker, signifying the formation of the new compound $Pd_xMoTe_2$. **b,** The same as **a**, but for 2H-$MoTe_2$. **c,** $R_{xx}$ of $Pd_xMoTe_2$ as a function of $T$, taken from D4 at various magnetic fields. **d**, d$V$/d$I$ map under varying $I$ and $T$. **e,** d$V$/d$I$ *v.s.* $I$ at 50 mK. A dip around ~ ±0.13 μA is seen in **e**, featuring a negative d$V$/d$I$ (note that this is not a negative resistance, but a differential resistance), which signifies a voltage perturbation due to superconducting transition in a neighboring region and hence inhomogeneity in this specific sample. **f-h,** The data set for 2D Pd metallized compound on monolayer 2H-$MoTe_2$ (D5).

Here, we focus on the transport properties of the new materials and find that in both cases they superconduct. Figs. 3c-e plot four-probe resistance measured on a Pd-metalized monolayer 1T'-$MoTe_2$, showing a $T_c$ ~ 0.45 K and a $B_c$ ~ 0.4 T. Similar values are observed in Pd-metalized monolayer 2H-$MoTe_2$ (Figs. 3f-h). The normal state resistance $R_n$ of these two devices is, however, quite different, being ~ 800 Ω for D4 whereas ~ 33 Ω for D5. This could be an indication that the resulting materials in the two cases may not be identical, although the transport device geometry plays a role. Consistent with the normal state resistance, the critical current in D4 ($I_c$ ~ 100 nA) is much smaller than that of D5 ($I_c$ ~ 1 μA), and consequently $I_cR_n$ does not differ by too much, consistent with the fact that $T_c$ is similar. It is possible that superconductivity resides on the Pd-Te layer formed in the structure. We note that in our high-resolution scanning transmission electron image of the $Pd_7WTe_2$ compound, no lattice structure can be identified as a single layer of known PdTe or $PdTe_2$ crystals [27]. Another possibility is that the superconductivity resides on the 2D Pd layer. Even though bulk Pd doesn't superconduct, the ultrathin Pd realized in our case has a unique lattice structure [27] and is possibly a superconductor. Also note that Pd hydrides superconduct, but this is unlikely the situation as our whole fabrication happens within Ar-filled glovebox. We don't have a conclusion on the exact atomic origin of superconductivity at this point but conclude that the new compound as whole is a superconductor.

## C. Designing superconductivity in a fractional Chern insulator

Recently, the fractional quantum anomalous Hall (FQAH) effect, a zero-magnetic field analog of fractional quantum Hall effect expected for fractional Chern insulators (FCIs), has been discovered in bilayer 2H-$MoTe_2$ twisted at an interlayer angle of 3° ~ 4° after a series of experiments at Seattle that uncovered its magnetism [36], Chern number [37] and the fractionally quantized Hall transport [38]. The thermodynamic evidence [39] and quantized Hall transport [40] of the FCIs in the same system have also been reported by two groups at Cornell and Shanghai, respectively. This is an exciting development in the field of topological and correlated phases of matter. One next question is to ask whether there will be interesting new phenomena if superconductivity is introduced to such systems. Theoretically, this could offer a possibility to realize new fractionalized electronics state, such as parafermion modes [2,6–9]. It is not yet known how to introduce superconductivity into this highly interesting but air-sensitive 2D material system. Conventional approaches based on deposition of elemental superconductors are difficult without reducing its quality. Here, we demonstrate that our on-chip 2D Pd metallization introduces superconductivity into twisted bilayer 2H-$MoTe_2$ in a designable fashion.

We fabricate twisted bilayer 2H-$MoTe_2$ (D6), at an angle of ~ 3.7° (determined using optical images during fabrication) that favors the FCI states upon electrostatic gating, in contact with pre-deposited Pd stripes which serve as both the Pd seeds and the electrodes for transport measurement (Fig. 4a). Figs. 4b & c show optical microscope images of the device before and after heat treatments at 220 °C for 40 mins, during which in-plane Pd transport occurs similarly to previous situations. Note that pristine monolayer and twisted bilayer 2H-$MoTe_2$ are insulators [see Supplemental Material Fig. S5 for characterization of the contact properties before and after a slight Pd transport]. With the Pd treatment, the resulting material turns into a metal that develops superconductivity with $T_c$ ~ 1 K and $B_c$ ~ 1 T, as characterized in Figs. 4d-g. $I_cR_n$ observed in this bilayer case is much larger than that of the monolayer seeded Pd compounds, indicating a larger superconducting gap, consistent with the higher $T_c$ and $B_c$. In the Supplemental Material Fig. S6, we include data taken from the same device but under in-plane magnetic fields, in which superconductivity can survive > 10 T, consistent the 2D nature of the superconductor.

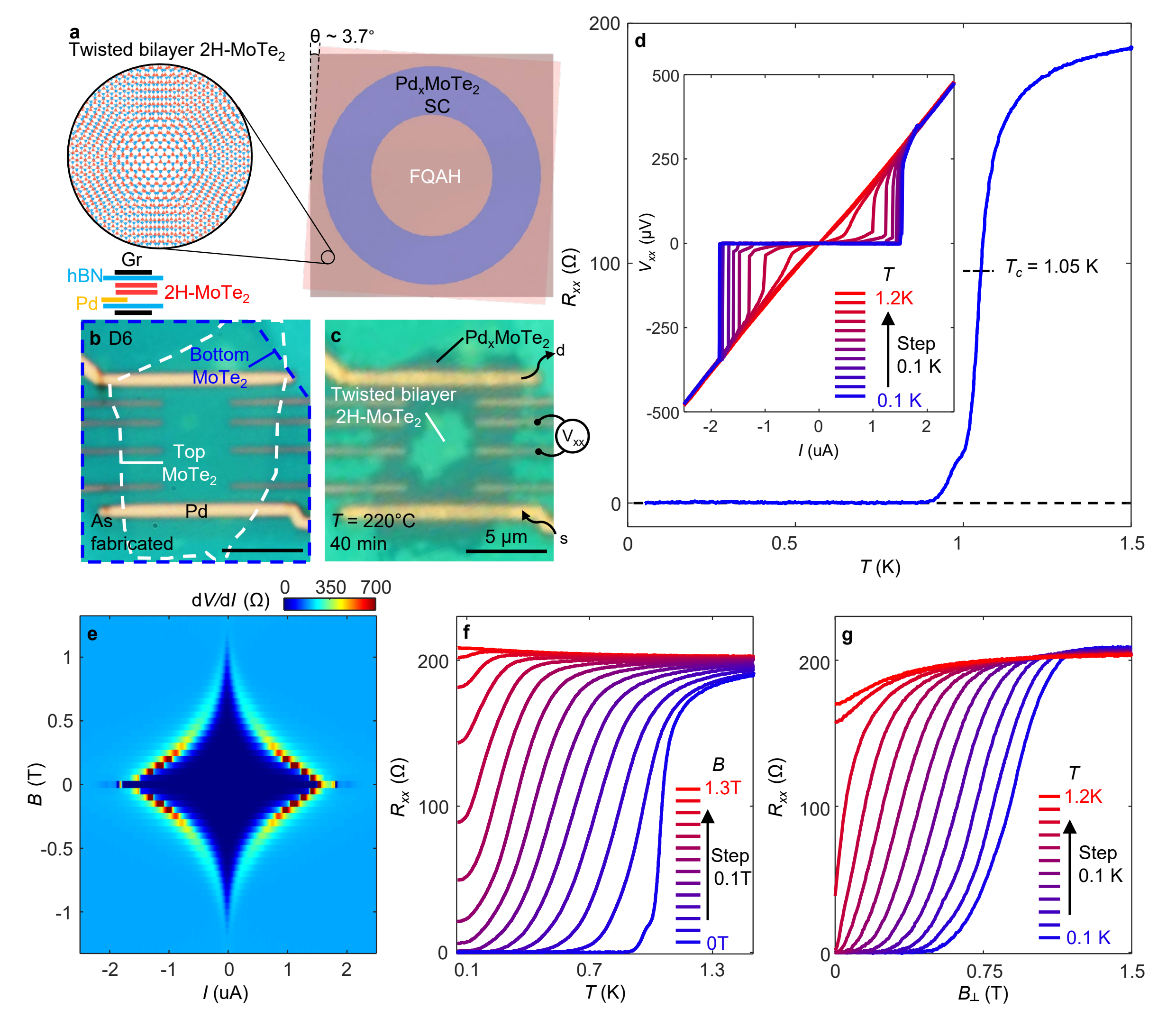

**Fig. 4. Designing superconductivity in a fractional Chern insulator. a,** Top: an illustration of an in-plane heterojunction of a superconductor and a fractional Chern insulator realized in twisted bilayer 2H-$MoTe_2$. The moiré lattice is shown to the left. **b** & **c,** Optical images of a device (D6) consists of Pd stripes (electrodes) and twisted bilayer 2H-$MoTe_2$, fully encapsulated with graphite/hBN stacks, before (**b**) and after (**c**) the heat treatment. **d,** $R_{xx}$ *v.s.* $T$ for $Pd_xMoTe_2$ realized in D6, revealing the superconducting transition. Inset shows the nonlinear *IV* curves taken at various $T$, displaying sharp jumps. The contact configuration used in the measurement is shown in **c**. **e,** A d$V$/d$I$ map taken under varying $B$ and $I$. **f,** $R_{xx}$ *v.s.* $T$ taken under various $B$. **g,** $R_{xx}$ *v.s.* $B$ taken at various $T$.

We further note that the resistance transition of our $Pd_xMTe_2$ superconductors typically occurs within ~ 0.2 K. This is much sharper than, for example, the superconducting transition in magic angle graphene [24], indicating a better homogeneity in our case. In some of our devices, a single $I_c$ peak is seen (e.g., Fig. 2e & Fig. 4e) but others develop multiple peaks (e.g., Fig. 3e & h), suggesting that inhomogeneity in different devices is different, as expected. We in general find that $Pd_xMTe_2$ grown on bilayer $MTe_2$ exhibits a better uniformity than that grown on monolayers.

Our results establish a feasible device fabrication approach to study the interplay between superconductivity and FCIs in a highly designable fashion. In this device (D6), we have already realized a loop-shaped superconductor (Fig. 4c), the inner of which resides the intrinsic twisted bilayer 2H-$MoTe_2$ that can be gated tuned into FCIs. The properties of a device interfacing an FCI and a superconductor in a lateral junction are of interest to the construction and search of non-abelian anyons. We envision fruitful future explorations along this direction based on the approach and device presented here as well as their variations.

## III. Discussion

Beyond Palladium, we have tested the phenomena on other metals. We don't find propagation of Au on $WTe_2$ at similar temperatures (Supplemental Material Fig. S7). We find Ni does propagate similarly on $WTe_2$, but the resulting new compound doesn't superconduct down to ~ 50 mK despite being metallic (Supplemental Material Fig. S8). The magnetic properties of the Ni-based compound however deserve further studies.

Our results establish a method of introducing superconductivity into a class of 2D topological chalcogenides. The rich topological phases and strong gate-tunability of the host 2D materials distinguish our approach from the previous attempts in proximitizing epitaxy-grown topological insulators. The size and shape of the superconducting islands can be controlled via designing the pattern of Pd seeds and manipulating the recipes of the heat treatment. One key feature is that the heat treatment only requires a temperature as low as ~ 200 $^{o}$C, which can be performed straightforwardly on a hot plate and/or under microscope. The whole process can be done inside a glovebox for devices on a chip without degrading the quality of sensitive components. We believe this unique approach for designing and creating robust superconductivity for air-sensitive 2D topological materials will enable a range of interesting explorations in condensed matter physics and superconducting quantum devices.

## ACKNOWLEDGEMENTS

This work is supported by AFOSR Young Investigator Award (FA9550-23-1-0140) to S.W. Electric transport measurement is partially supported by NSF through the Materials Research Science and Engineering Center (MRSEC) program of the National Science Foundation (DMR-2011750) through support to L. M. S. and S.W. and a CAREER award (DMR-1942942) to S.W. Device fabrication is partially supported by ONR through a Young Investigator Award (N00014-21-1-2804) to S.W. S.W. and L.M.S. acknowledge support from the Eric and Wendy Schmidt Transformative Technology Fund at Princeton. S.W. acknowledges support from the Sloan Foundation. L.M.S. acknowledges support from the Gordon and Betty Moore Foundation through Grants GBMF9064 and the David and Lucile Packard Foundation and the Sloan Foundation. Y.J. acknowledges support from the Princeton Charlotte Elizabeth Procter Fellowship program. T.S. acknowledges support from the Princeton Physics Dicke Fellowship program. A.J.U acknowledges support from the Rothschild Foundation and the Zuckerman Foundation. K.W. and T.T. acknowledge support from the JSPS KAKENHI (Grant Numbers 21H05233 and 23H02052) and World Premier

International Research Center Initiative (WPI), MEXT, Japan.

Supplemental Materials for

# Superconductivity from On-Chip Metallization on 2D Topological Chalcogenides

Yanyu Jia[1,*], Guo Yu[1,2], Tiancheng Song[1], Fang Yuan[3], Ayelet J Uzan[1], Yue Tang[1], Pengjie Wang[1], Ratnadwip Singha[3], Michael Onyszczak[1], Zhaoyi Joy Zheng[1,2], Kenji Watanabe[4], Takashi Taniguchi[5], Leslie M Schoop[3], Sanfeng Wu[1,*]

[1] *Department of Physics, Princeton University, Princeton, New Jersey 08544, USA*
[2] *Department of Electrical and Computer Engineering, Princeton University, Princeton, New Jersey 08544, USA*
[3] *Department of Chemistry, Princeton University, Princeton, New Jersey 08544, USA*
[4] *Research Center for Electronic and Optical Materials, National Institute for Materials Science, 1-1 Namiki, Tsukuba 305-0044, Japan*
[5] *Research Center for Materials Nanoarchitectonics, National Institute for Materials Science, 1-1 Namiki, Tsukuba 305-0044, Japan*

*Email: sanfengw@princeton.edu; yanyuj@princeton.edu

**This file includes**

Device fabrication and transport measurement

Figures S1 to S8

Reference [1–3]

## Device Fabrication

hBN and graphite flakes were exfoliated on $SiO_2$/Si substrates, identified and characterized under an optical microscope and atomic force microscope (AFM, Bruker Dimension Edge or Bruker Dimension Icon). hBN flakes were either transferred directly to $SiO_2$/Si substrates (D1, D4, D5, D7, D8), stacked on prepatterned metal gate (D2, D3) on those substrates, or stacked with graphite gate (D6) before transfer to $SiO_2$/Si substrates. Electron beam lithography, followed by cold development, reactive ion etching, and metal deposition, are then applied to deposit the patterned Pd (or Au, Ni) seed sources (typically 10 ~ 20 nm thick) on the hBN flakes. After metal deposition, the bottom stacks were then tip-cleaned using AFM under the contact mode. We exfoliated monolayers and bilayers of $T_d$ -$WTe_2$, 1T'-$MoTe_2$ and 2H-$MoTe_2$ in an Ar-filled glovebox. Right after a suitable TMD flake is identified, they were picked up by another hBN flake (sometimes with a graphite fake at the top) and then stacked onto the bottom patterned with Pd seeds. For transport devices, some of the Pd seeds were connected to large metal pads outside the vdW stack for wire bonding. The whole process involving $WTe_2$ and $MoTe_2$ was performed in glovebox with $H_2O$ < 0.1 ppm and $O_2$ < 0.1 ppm. The vdW devices prepared above were placed on a hot plate, with controlled temperature for Pd metallization. The process was carefully monitored under an optical microscope. Detailed characterization of the in-plane mass transport and 2D chemical process and the atomic structure of the resulting new compounds, especially $Pd_7WTe_2$, can be found in our previous report [1].

## Transport Measurements

We performed electrical transport measurements on the devices in a dilution refrigerator equipped with a superconducting magnet and a base temperature of ~ 20 mK (the electron temperature ~ 32 mK). Four-probe resistance measurements were performed using the standard ac lock-in technique with a low frequency, typically ~ 13 Hz, and an ac current excitation. A dc current was also applied to the source electrode for critical current measurements.

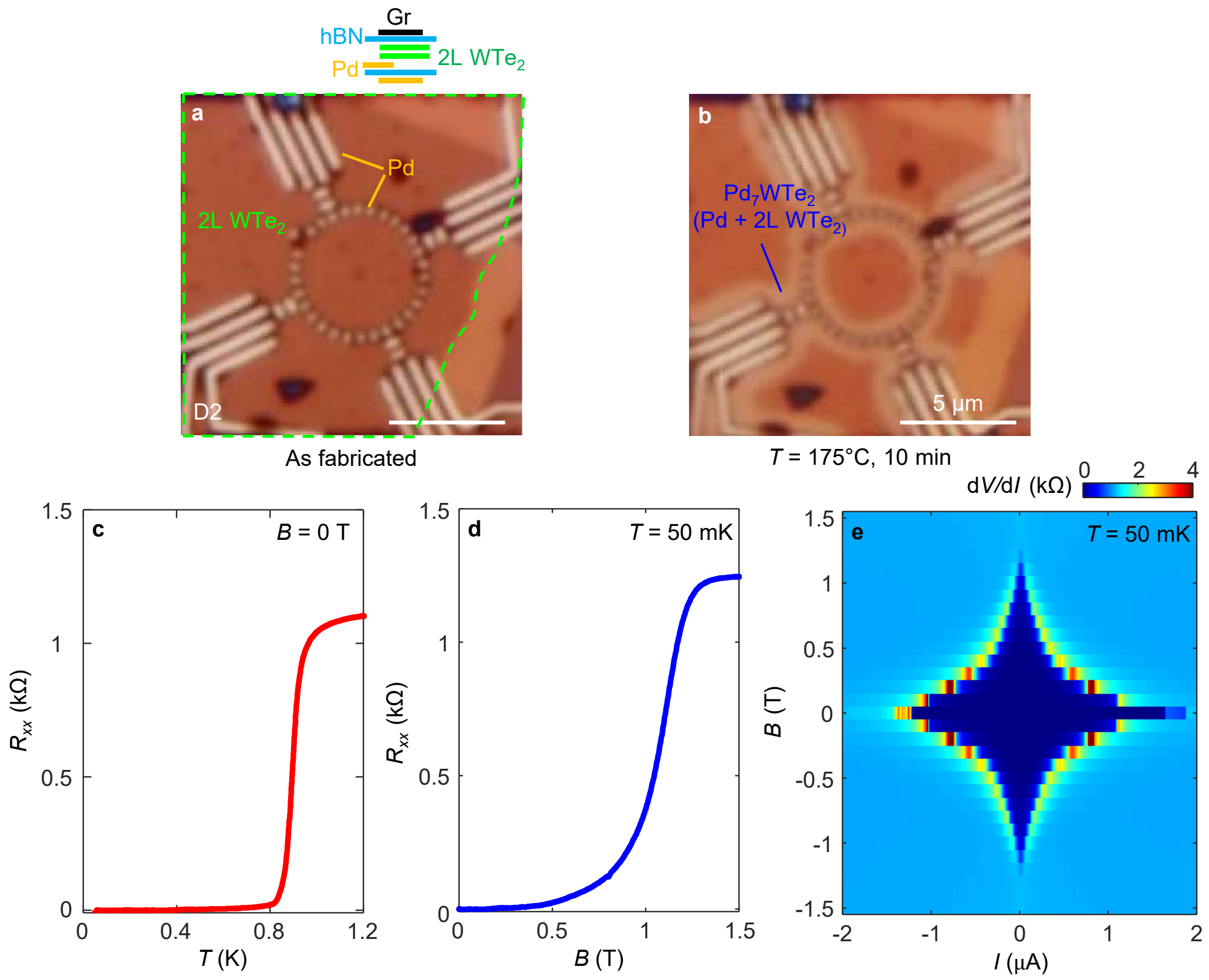

**Fig. S1. Superconductivity in $Pd_7WTe_2$ grown from Pd metallization on bilayer $WTe_2$. a & b,** Optical images of the device (D2) before and after the heat treatment. Bilayer $WTe_2$ covers the region outlined by the green dashed line. The Pd seeds are purposely designed into a ring shape to illustrate one way to control the superconducting geometry. Pd stripes also serve as electrodes. Inset on the top illustrates the device structure in the cross section. **c & d,** Resistance as a function of temperature (**c**) and magnetic field (**d**), clearly revealing the superconducting transition of $Pd_7WTe_2$ in this device. **e,** d$V$/d$I$ map under varying the dc source current $I$ and magnetic field $B$.

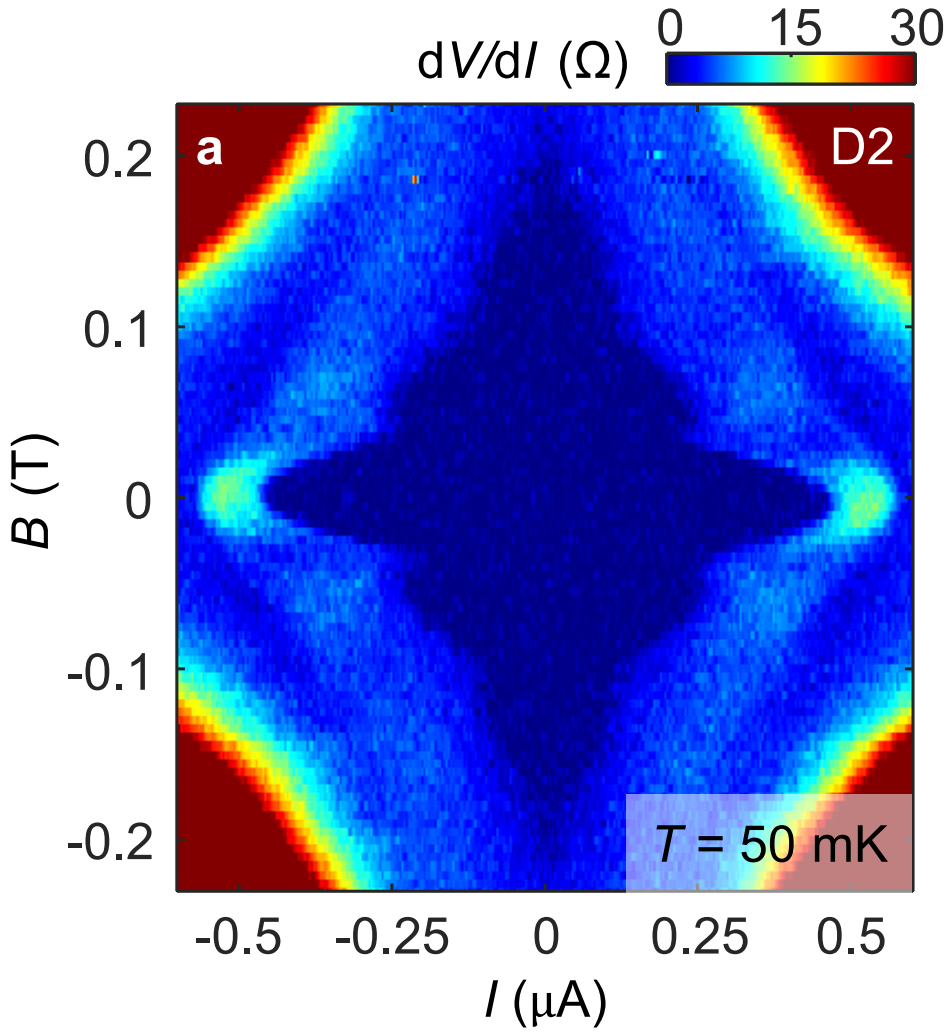

**Fig. S2. Fraunhofer-like pattern. a,** Differential resistance d$V$/d$I$ versus bias current $I$ and perpendicular field $B$. Fraunhofer-like interference pattern is observed, indicating an accidental formation of a Josephson junction arising from disorder and the phase-coherent transport of superconductivity in D2.

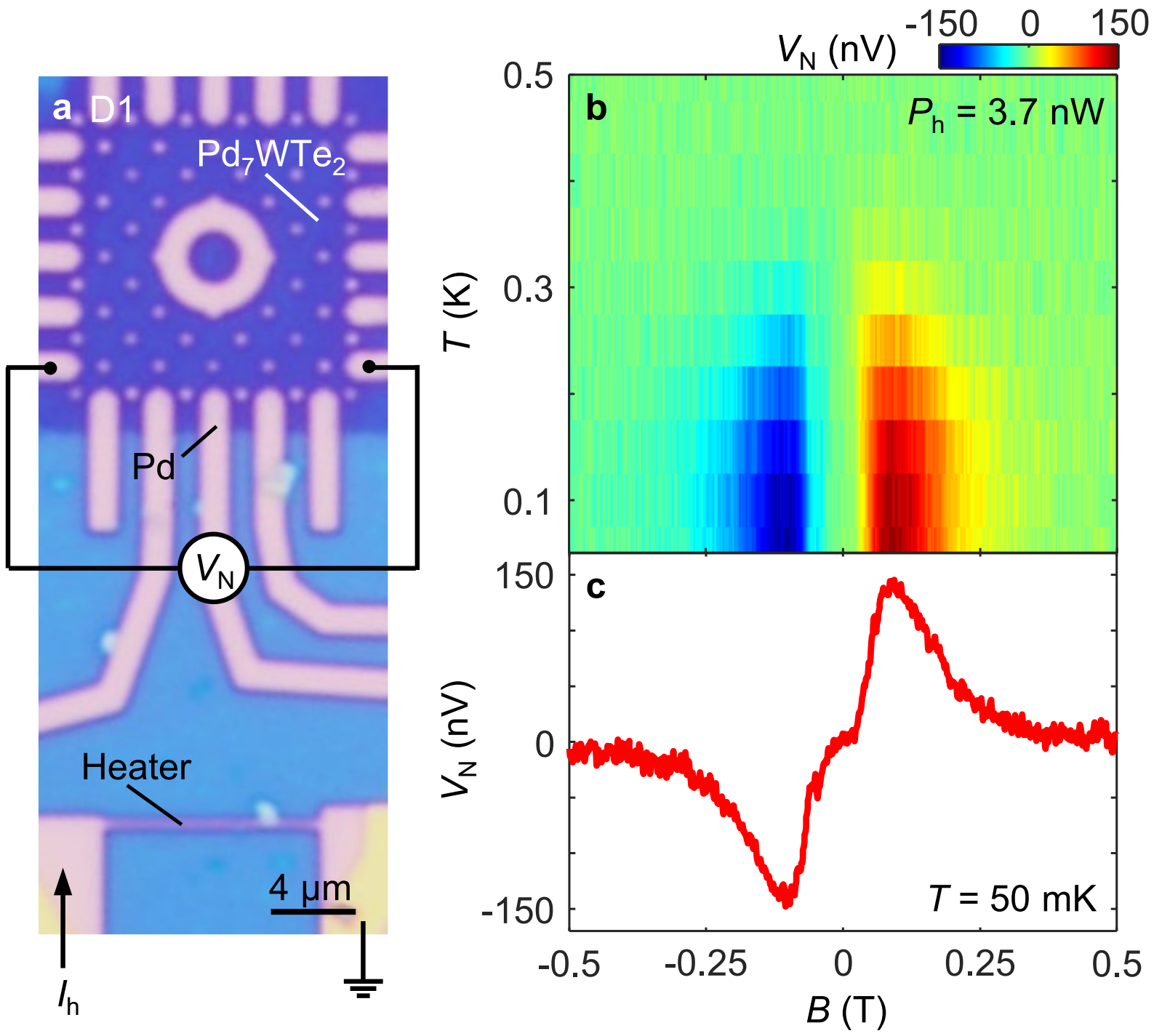

**Fig. S3. Vortex Nernst effect in $Pd_7WTe_2$. a,** An optical image of device D1 after Pd metallization on monolayer $WTe_2$. The contact configuration for the Nernst experiment is indicated. A current is applied through the heater to produce a temperature gradient on the sample, which drives the motion of superconducting vortices (under $B$) across the $Pd_7WTe_2$. Detailed discussion about Nernst measurements can be found in ref [2,3]. **b,** Temperature dependence of Nernst voltage ($V_N$) *v.s* $B$. **c,** $V_N$ *v.s* $B$, taken at $T$ = 50 mK. The observed vortex Nernst effect signifies the formation and motion of superconducting vertices.

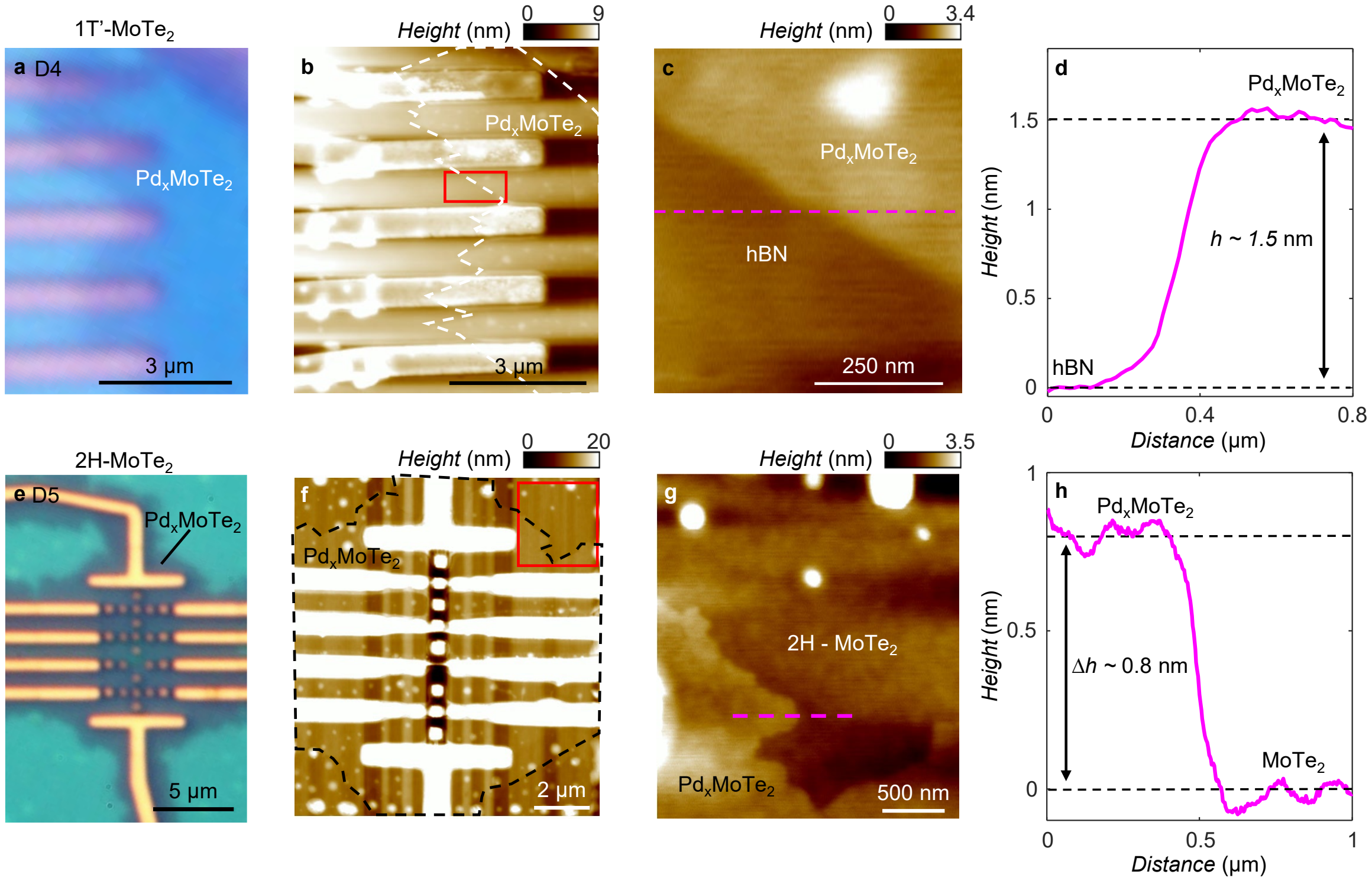

**Fig. S4. AFM analysis of D4 & D5. a,** The optical image of the device (D4) after heat treatment. **b,** An AFM image of the device, where the $Pd_xMoTe_2$ formed on the originally monolayer 1T'-$MoTe_2$ region is outlined by the white dashed line. **c**, A zoomed-in AFM image of the region indicated by the red rectangle in **b**. **d,** The height profile along the line cut shown in **c**, showing a height of ~ 1.5 nm for $Pd_xMoTe_2$ in this device. **e-h**, The same AFM analysis on device D5 ($Pd_xMoTe_2$ grown on monolayer 2H-$MoTe_2$), revealing a consistent height of ~ 0.8 nm Pd layer (**h**) and therefore ~ 1.5 nm for the thickness of $Pd_xMoTe_2$. Note that this height is very close to the measured AFM thickness of $Pd_7WTe_2$ grown on monolayer $WTe_2$ (see [1]), indicating that a similar amount of Pd is formed here on $MoTe_2$ and thus $x$ ~ 7.

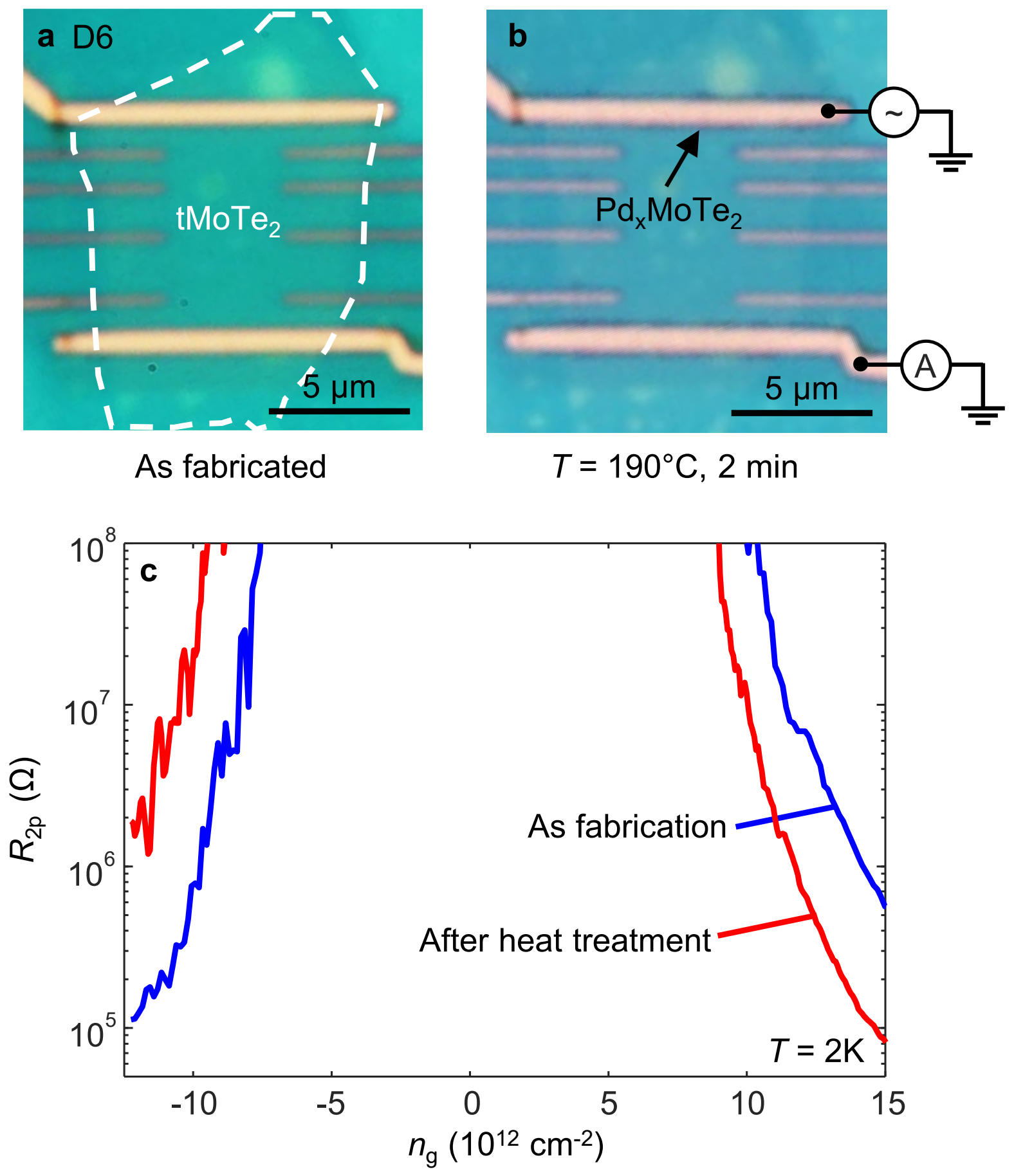

**Fig. S5. Contact properties of a $tMoTe_2$ device, before and after slight Pd metallization. a & b,** Optical images of D6 before (**a**) and after (**b**) slight Pd metallization, indicated by the narrow darker area surrounding the Pd contacts. **c,** The gate induced doping ($n_g$) dependence of two-probe resistance ($R_{2p}$) for device D6 as fabricated (blue) and after annealing (red).

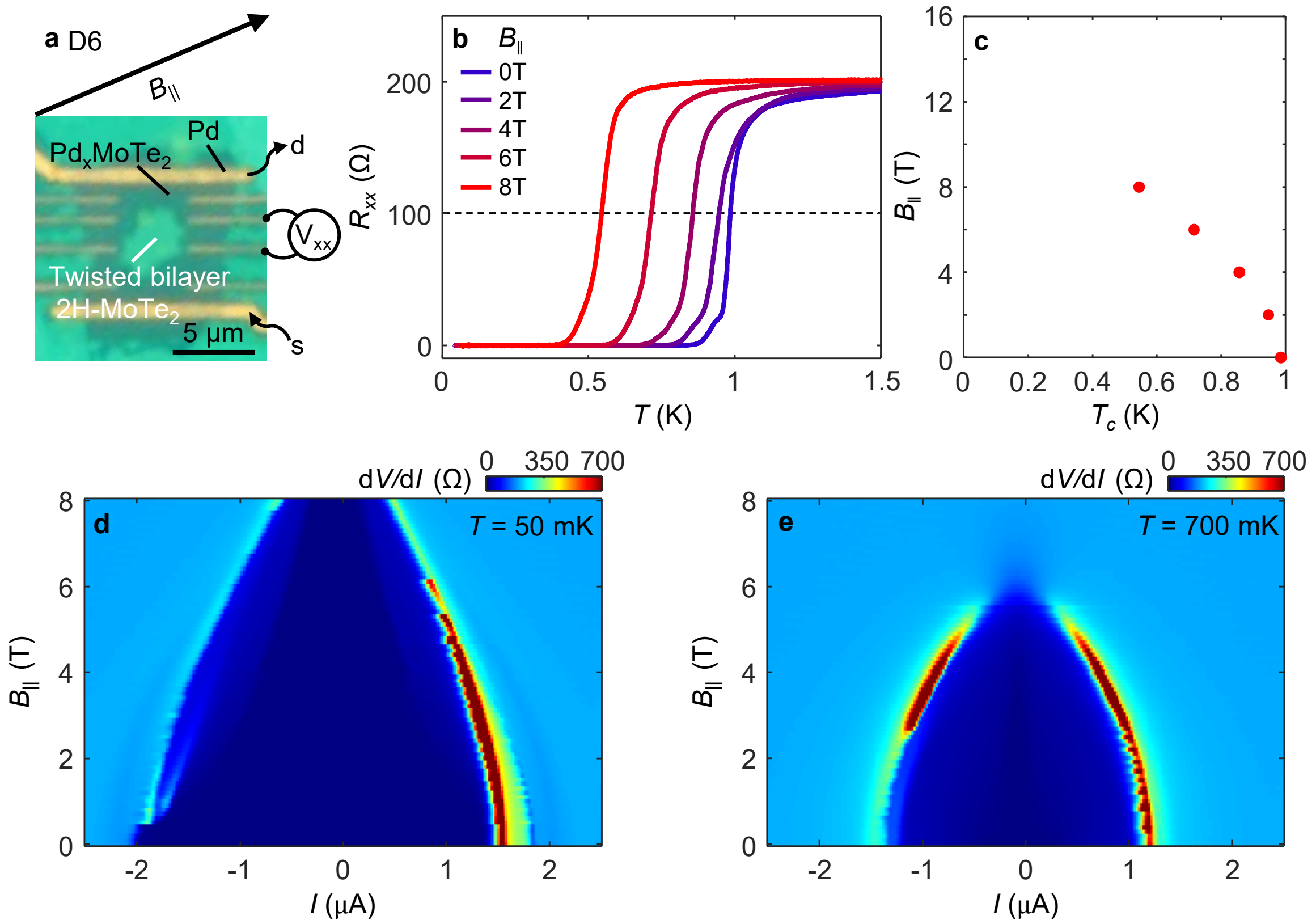

**Fig. S6. In-plane magnetic field dependence of superconductivity in Pd-metallized $tMoTe_2$.** **a,** An optical image of the device (D6) after Pd metallization on $tMoTe_2$. The measurement configuration and in-plane magnetic field direction are indicated. **b,** $R_{xx}$ *v.s.* $T$ under varying the in-plane magnetic field ($B_{\|}$) up to 8T (the highest field in our setup). **c,** The extracted $T_c$ *v.s.* $B_{\|}$. The critical $B_{\|}$ at the base temperature is expected to be > 10 T. **d,** A d$V$/d$I$ map taken under varying $B_{\|}$ and $I$ at $T$ = 50 mK. **e,** The same d$V$/d$I$ map at $T$ = 700 mK.

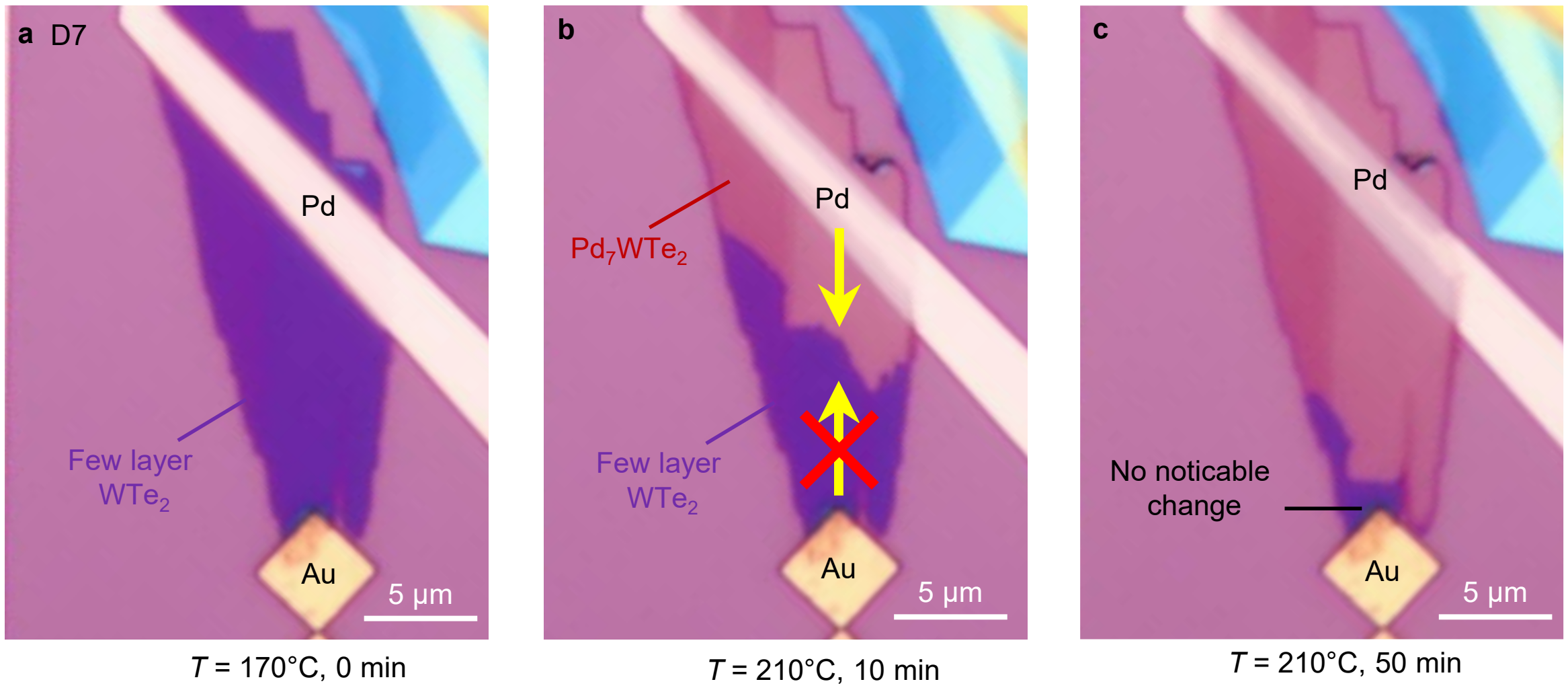

**Fig. S7. Contrast behaviors between Pd and Au. a,** An optical image of the device (D7) as fabricated. During fabrication, the temperature is kept below ~ 170 ºC. A few layer $WTe_2$ is placed in contact with pre-deposited Pd and Au seeds. No mass transport is seen under an optical microscope. **b,** An optical image of the device after heat treatment at 210 ºC for 10 min, clearly revealing the Pd transport and reactions. No Au transport is seen under the same condition. **c,** The optical image after ~ 50 min.

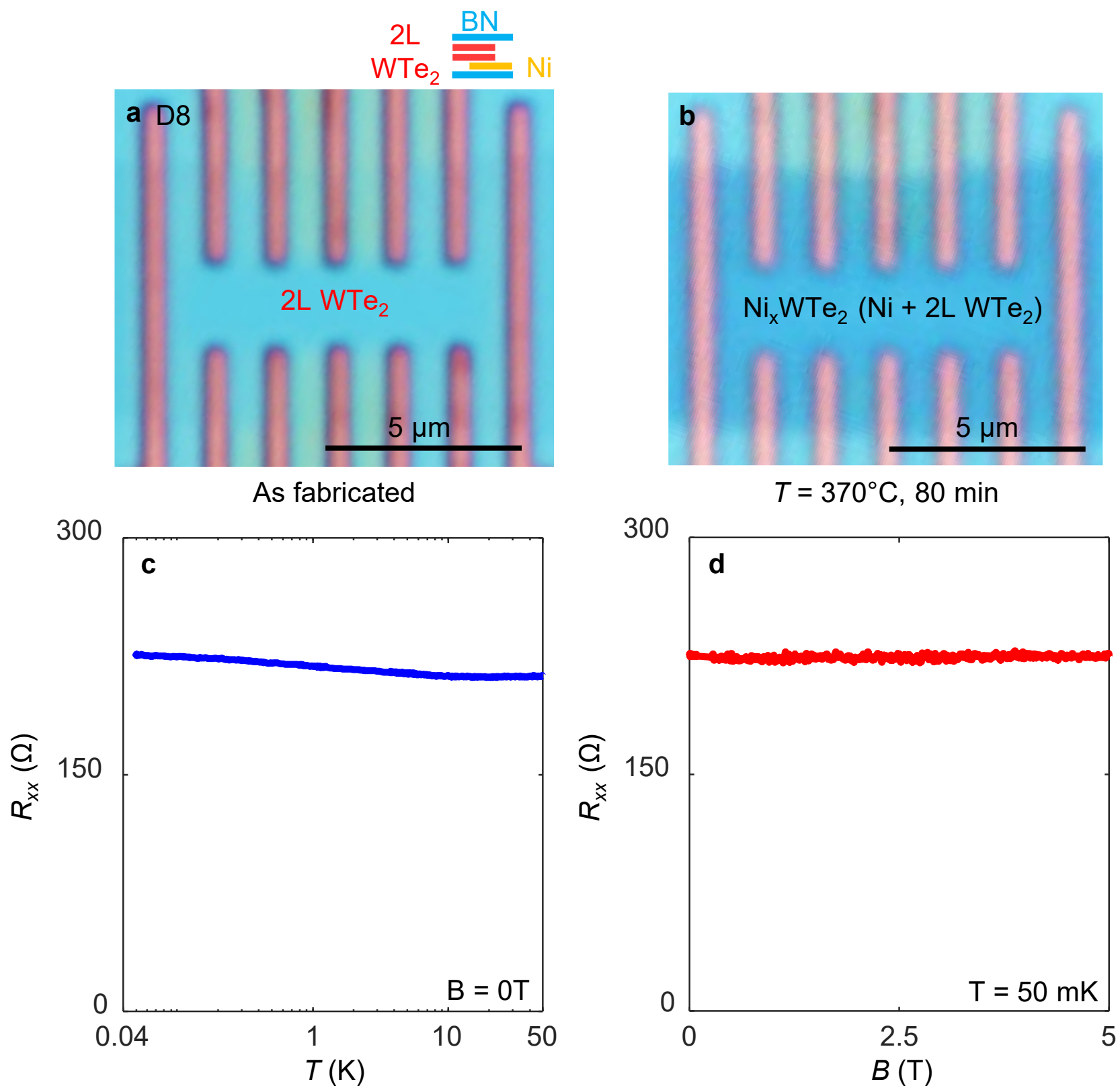

**Fig. S8. Ni transport on bilayer $WTe_2$ and the absence of superconductivity. a,** An optical image of the device (D8) consisting of Ni stripe (electrodes) and bilayer $WTe_2$ encapsulated by hBN, as fabricated. **b**, An optical image of the device after heat treatment at 370 °C for ~ 80 mins. Ni fully covers the $WTe_2$ region under the view, as indicated by the darker color. **c,** Four probe resistance $R_{xx}$ of $Ni_xWTe_2$ measured in the device, as a function of temperature (**c**) and magnetic field (**d**). No superconductivity is found down to 50 mK.